\documentclass[a4paper]{jpconf}
\usepackage{braket}
\usepackage{graphicx}
\usepackage{amsfonts}
\begin{document}

\title{Quantumness, Randomness and Computability}
\date{}

\author{Aldo Solis,  Jorge G. Hirsch}

\address{Instituto de Ciencias Nucleares, Universidad Nacional
Aut\'onoma de M\'exico, Apdo. Postal 70-543, M\'exico 04510 D.F.}

\ead{hirsch@nucleares.unam.mx}

\begin{abstract}
Randomness plays a central rol in the quantum mechanical description of our interactions.
We review the relationship between the violation of Bell inequalities, non signaling and randomness.
We discuss the challenge in defining a random string, and show that algorithmic information theory provides a 
necessary condition for randomness using Borel normality.  We close with a view on incomputablity and its implications 
in physics.

\end{abstract}


\section{Introduction}
The empirical success of quantum mechanics in the description of microscopic phenomena is indeed impressive.
We visualize matter as made of molecules, formed by atoms with electrons orbiting around a nucleus with protons
and neutrons, which are built with quarks, bound together by interaction carriers, like gluons, pions and photons.  
The mathematical formalism allows the description of element formation inside the stars, the structure of materials, 
the energetic collisions which take place in many scenarios including high-energy accelerators. A plethora of new technologies, 
devices, materials, has born out of it. 

At the same time, quantum physics has put at the reach of our hands the limitations of our imagination, of our capacity to describe
employing words this microscopic world. The very notion of reality is waning, despite many efforts on the opposite direction.
Particles and waves are two central notions of the classical description of the physical world. They are associated with the objects, 
formed by particles, and the interactions between them, whose perturbations propagate as waves. Particles are localized, waves
have extended fronts. A century of wave-particle duality has messed it all. The violation of the Bell inequalities opened a Pandora's box. If we insist in having well defined objects in the quantum domain, they must have very strange properties: a measurement of an observable property in one part of a quantum system is correlated with the outcome of a different measurement realized in other part of the system in a way which can be described as non-local, contextual and at random. This property, called entanglement, is found in most composed quantum systems. Good overviews on these subjects were published associated with the 50 years of Bell's theorem \cite{bell,wiseman14}.

By non-local it is implied that the correlations in the outcomes of the measurements are obtained in space-time regions which are
causally disconnected, i.e. no signal traveling at most at the velocity of light in vacuum can be sent from one region to the other. 
Contextual refers to a connection between the measurement device and the property to be measured. The quantum formalism  
associates observables  with expectation values of non-commuting operators, implying the uncertainty principle. The better the value of one observable is known, the lesser we can say about the other. More than that, if these complementary properties are assumed to have definite but unknown values when they are not measured, flat contradictions appear in their description.  The simplest way to avoid these contradictions is to assume that, if these properties exist at all, they manifest themselves in accordance to the measurement device, the context in which the experiment is performed.    

If the quantum correlations are associated with quantum objects, they must have non-local and contextual properties. What would prevent us to employ contextuality, measuring in a given way one part of  quantum system to generate through the non-local correlation a specific outcome of the measurement in the other part, far away,  sending in this way a signal faster than light? We cannot do that, because we can select the property to be measured, but cannot influence the outcome: it is either trivial, when we measure a property which was already prepared to have a specific value, or random, when we measure a complementary property.

We can see that randomness is an imprescindible companion of non-locality and contextuality, if no-signaling (the short name of the impossibility to send information faster than light in vacuum) would be enforced. And we really want it, because special relativity tells us that being able to send information between causally disconnected space-time regions is equivalent to send information back in time is some reference frames. And this would cause logical paradoxes which would demolish the basics of science as we know them.  

When instead of thinking about objects and their interactions, the quantum system is analyzed from the information perspective, interesting alternative views arise.  Reality and information can be considered as two sides of the same coin. Randomness,
complementarity and entanglement emerge from the fact that from individual measurements there is a finite amount
of information available. One way to formalize this ideas is to postulate that an elementary system can only give a definite result in one specific measurement. It follows that the outcomes of other independent measurements must  be irreducibly random \cite{Zei99}.

When the intrinsic quality of quantum randomness is accepted as a postulate, a world of applications emerges. 
``Private randomness" is defined by the presence of correlations that cannot be reproduced with local variables. It is quantified by the violation of Bell inequalities, and is associated with the impossibility to predict a given string employing a classical computer and classical information \cite{Sca10,Pir10,chris13,bancal14,nieto14}. 

The concept of probability is associated with quantum predictions through the Born rule: the square of the wave function provides the probability density of observing a given event. It is associated with an intrinsic quantum randomness. Employing composite systems, like pairs of entangled photons, the violation of Bell inequalities guarantees their quantum origin. It is assumed to imply that the information generated with their measurements did no exist before them, certifying their privacy, and that they are ``as random as they can be". 

Can we prove that a sequence of measurements is already random? This is a challenging question, whose answer is explored in the following section.   

\section{Randomness}

How can be know that a given series of numbers is random?
Suppose we give you the sequence ``1,5,7,0,7''.  It looks pretty random,
and maybe you accept it as a good candidate for a sequence of random numbers: 
But if we give you the numbers ``1,2,3,4,5,6'' instead; you can say:
``Well, that's not random, you are just counting''. 
We offer you 
the sequence ``2,4,6,8,1,0,1,2''; and you answer: ``you are saying the same
numbers times two!''.
With our 
best effort 
we come
with the
sequence ``1,1,2,3,5,8'' and irritated you say: ``That's the beginning of
the Fibonacci sequence!''. 
Even if you accepted the first sequence as random, we may say it is not, a closer look reveals that 
it contains the first digits
of $\pi/2$.

It looks like we can always find a way to argue that a sequence is not
random. The sequences above are ``easy'', but take  ``8,3,2,0,9"; it does not
seem to have a pattern and if we say something like: take the first
digit and add -5, then add -1, then -2 and finally add 9, it may look
like we just made that up. So what descriptions can be considered as
patterns? 

The notion of randomness in this context is ``the absence of a pattern".
%
The idea of pattern refers to a procedure that describes the string. Such 
a procedure doesn't need any kind of decision or creativity, is just a set
of instructions that need to be followed to reproduce the string in an
``algorithmic'' way \cite{chaitin}.

\subsection{Turing Machines}

One of the greatest achievements in the past century was the formalization
of the algorithmic procedures, this was done in many different ways and all
of them turn out to be equivalent. The most intuitive one is due to Alan M. 
Turing and is called ``Turing Machines''\cite{art:turing}.
	An excellent exposure of this subject can be find in \cite{davis}.

A Turing machine can be seen as a man following a set of rules, this man 
can read and write in a boundless
(we don't want to limit the amount of paper he can use)
tape divided in squares, every square can have only one symbol at a time.
He can read only one symbol at a time
and, depending on that symbol and the current state of mind (``memory and thoughts'')
he will write another symbol, change his state of mind and move to the
left or right.%
\footnote{
	The way to formalize memory and thoughts is just by defining
	a finite set of states $\{q_i\}$. Every time the machine makes an 
	operation the state of the machine changes to another state $q_j$.
	The concept of state of mind can be applied to machines too. In
	the case of a calculator made with gears this states describe the
	physical state of the gears.
}

This definition may seem abstract. However, nowadays we are so used to 
manipulate computers that another way to make this a comprehensible definition is to
associate a Turing machine with a general purpose computer which has an infinite amount
of memory (because the tape is boundless), running a specific program. 
This brings an interesting question: Is there a Turing machine that can be identified
with a general purpose computer running any program? Yes, this is a really 
important class of
machines called Universal Turing Machines (UTM). Every UTM can ``run'' every
possible program, i.e. can mimic the behavior of every other Turing machine.
This is done with the proper input, in the same way a general purpose computer
can ``run'' every algorithm, which must be written in an appropriate code. Therefore, for 
every machine $M$ with input $i$ must exist an input $x$ such that a UTM $U(x)$
can imitate $M(i)$ when is feed with $x$, and therefore $M(i)=U(x)$.

\subsection{Information, Randomness and Turing machines.}

The existence of a UTM is very important for our purposes.
We cannot take
a Turing machine as a description of a pattern, because for
every finite sequence we can make a Turing machine whose output is the
sequence. Just take a program that says something like 
$$
print \,\,``0111001..." 
$$
(putting the sequence as argument of 'print').
In this way
we can always find an algorithm that prints the sequence. 
The trick is simple:
the sequence is contained in the program. But then, the number of symbols
 needed to print the sequence is by necessity larger than the sequence itself.
 
 We can make contact with the intuition suggested above:
when  a pattern gets really complicated we think that it should not be considered as
a pattern.
In the realm of algorithms we can say: if the program is longer than the
sequence itself that's not a pattern. This gives us the opportunity to define
a pattern in a sequence as a program, whose output is the sequence, with fewer
symbols than the sequence. The sequences without patterns as those we want to call random.

Our relation with  computers provides us with a familiar example, where 
the key concept is compression. 
We employ many compression algorithms, like zip,
to save space in the hard drive or send information faster. 
We can identify
the strings that does have patterns with the compressible ones, because there is
an algorithm whose output is the string.
It offers a very good intuition about
random strings: A random string cannot be compressed. 

It may seem hard to find a random string but that's not the case.
As an example take all the binary strings
with 3 symbols, there are 8 of them. And there are only 6 strings shorter:
2 binary strings with one symbol (0 and 1), and 4 with two symbols (00, 01, 10 and 11).
If we try to represent the binary strings with 3 symbols with shorter strings, at least two of them
cannot be included,  therefore there are at least 2 random 
strings. Algorithms are programs in a given language, which are represented by binary strings also. 
This argument can be applied to any number of symbols, implying that
random strings are quite common. 

\subsection{Algorithmic information theory}
 
The algorithmic information theory
	\cite{chaitin,book:Calude,book:Li}
was developed
in the past century 
by Andrey Kolmogorov, Ray Solomonoff, Gregory Chaitin and others. 
Within this
theoretical framework 
we can,
instead of just talking about random and non-random strings,
ask: what is the smallest number of
symbols necessary to print a given sequence? 
For example, take the sequence made of '1'
a million times. This sequence have a million symbols but it will be very simple 
to write a program to print it, and such a program will easily have less than 
a million symbols.

The least number of symbols is called the algorithmic information content of the
sequence. Random sequences have, by definition, an information content close to their length.
Other mathematical objects, like real numbers, 
can be identified with a binary sequences using their binary expansion. In this way
we can define the algorithmic information in a real number.
It is worth to mention that the paper in which Turing introduce Turing machines is dedicated 
	to computable numbers, showing the most real numbers are incomputable.

With this formalism we can investigate randomness, in the way we have just defined it.
Can we use it to construct an algorithm able to classify strings in random and non-random?
Unfortunately, this is not possible. The demonstration is simple.

Suppose we have a method to distinguish between random and non-random strings, 
and this method is algorithmic, so can be written as a program and feed to a UTM.
To complement it, we write another piece of code whose output contains all the possible strings, 
one by one, in order, i.e. 0,1,00,01,10,00,... . With these programs we can make
another program following these steps:
 
\begin{enumerate}
\item produce a sequence,
\item check if the sequence is random,
\item if it is random print it,
\item go to (i).
\end{enumerate} 

The outcome of these instructions is a sequence of random strings, all of them
in order of length. Suppose the length of this new program is $L$ symbols,
this program will eventually print a random string of $L+1$ symbols. But then we
have just printed a random sequence of $L+1$ symbols with less than $L+1$ symbols!
This contradicts the definition of algorithmic randomness, therefore there cannot exist an algorithm
that can decide the randomness of a string.

\subsection{Prefix free strings and the Omega number}

\begin{figure}
\centering
\includegraphics[width=0.7\textwidth]{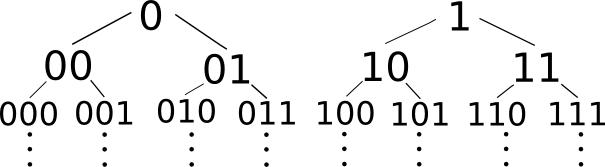}
\caption{All possible binary string ordered in a tree structure.}
\label{arbol}
\end{figure}

The algorithmic information theory have changed over time and today is really
common to find it as prefix-free algorithmic information theory, where prefix-free is
referred to the possible programs accepted by the UTM.
For instance, if the UTM
accepts ``1000'' as a program, then the string ``10001'', or any other starting with ``1000'', 
can't be accepted. In the figure \ref{arbol} we represent all the possible binary strings, showing explicitly
the first three lines, containing the sequences with 1, 2 and 3 symbols, respectively.
Below every string we find all the possible extensions of this string.
If a string is accepted as a program, all the possible strings below it can't be accepted as programs.

A schematic procedure to obtain the accepted programs could be this:
we have a coin and want to produce a program by flipping it several times,
following these steps:
\begin{enumerate}
\item Flip a coin.
\item Incorporate the output (0 or 1) at the end of the string you already have.
\item Check if the string of results is an accepted program.
\item if it is accepted, halt; if not go to (i).
\end{enumerate}

Note that we are not missing any possible program, because every time we obtain an
accepted program, it closes a branch of the tree in Fig. 1. We are not allowed to continue
adding new symbols to this string (because of the prefix-free condition), and must start 
building a new string from its first symbol. For those of us who are familiar with programming,
this step is like compiling the code in a given programming language, and checking if it compiles,
i.e. is an accepted program, or not. 
Each different string of length $n$  has a probability $P_i = \frac{1}{2^n}$ of being generated. The prefix-free condition
guarantees that  $ \sum_{i= \mbox{\tiny possible programs}} P(i)= 1. $

Once we have an accepted program, which in this view includes all the input data files, the next question is:
will it generate an output? Which is equivalent to the question: will this program stop, or will it be running forever?
Running a program we can learn if it halts, in the event it halts. The problem is to decide when to give up on a program that does not halt. While many special cases can be answered, Turing demonstrated that a general solution is impossible. No algorithm, no mathematical theory, can ever tell us which programs will halt. The demonstration goes in the following direction:
Suppose there is an algorithm that takes the description of a Turing machine and some input x, this algorithm does not halt if the machine halt when is feed with the input x and halt if the machine 
will be running forever.
Because this is an algorithm there must be a Turing machine that formalize this, and we can feed this algorithm with its own description, and because of the way we define it 
this machine does not halt if it halts, which is an obvious contradiction. Therefore there cannot exist an algorithm able to decide if a Turing machine will halt. This is known as the halting problem, and is closely related with G\"odel's theorem, which states that there is an infinite number of true statements in Mathematics which cannot be formally proved \cite{chaitin}.

Chaitin considered the ensamble of all possible programs, and asked the question:  What is the probability of getting a halting program choosed at random from the ensamble, as outlined  in the example given above? 
If we can associate to each accepted program $i$ a function defined as $H(i)= 1$ is the program halts, and $H(i)= 0$ if not,
we can evaluate the probability of obtaining a halting program for a given Universal Turing Machine U as
%
$$
	\Omega_U=\sum_{i= \mbox{\tiny possible programs}}H(i)P(i)=\sum_{i= \mbox{\tiny possible programs}}H(i)\frac{1}{2^{n(i)}}
$$
where $n(i) $ is the number of bits (symbols in the general case) in the program $i$. $\Omega_U$ is called the 
Chaitin's Omega number for U.

Omega is perfectly well defined. It is a specific number which can have a value between zero and one (is a probability), but it is impossible to compute in its entirety. Some of its first digits can be calculated. For example, if the computer programs 1, 00 and 010 halt in a given universal Turing Machine, the first digits of this $\Omega_U$ would be 0.111.  But the relevant aspect is that the first N digits of $\Omega_U$ cannot be computed using a program significantly shorter than N bits long. In this sense, any $\Omega_U$ has an infinite number of bits which are well and uniquely defined, but cannot be calculated by any finite program \cite{chaitin}. It does not matter how long is the program, the number of bits which are impossible to calculate is infinite. 
It is worth to mention that $\Omega_U$ is completely different from some irrational numbers, like $\pi$ or $\sqrt{2}$, for whom billions of digits can be computed with quite compact programs \cite{pi}.

The consequences of the impossibility to compute $\Omega_U$ go far beyond computational sciences. Together with the findings of G\"odel and Turing, they demonstrate that, given any finite set of axioms, there is an infinite number of truths that are unprovable in that system. It implies directly that an infinite number of mathematical true statements are impossible to prove, they must be assumed as new axioms, without any justification.  Philosophically, we are touching the limits of the principle of sufficient reason, which states that everything happens for a reason or, in other words, if something is true, there must be a reason for that. In mathematics most of what is true has no reason at all \cite{chaitin}. 

\section{Randomness in physics and Borel normality}

The close connection between the previous section and the foundation of physics can be seen in the following quote
``The discovery that individual events are irreducibly random is probably one of the
most significant findings of the twentieth century. ... for the individual event in quantum
physics, not only do we not know the cause, there is no cause. The instant when a
radioactive atom decays, or the path taken by a photon behind a half-silvered beamsplitter
are objectively random "\cite{art:Zei05}.

In physics we 
employ the concept of random in different contexts.
Sometimes we say something is random when we ignore the state of the system,
sometimes when the behavior of the system is really complicated (that's
really close to the previous definitions), and finally we have intrinsic randomness, applied to 
individual quantum systems where we can't, in general,  predict the
result of individual measurement outcomes.

What we mean when we say something is unpredictable? In the case of 
quantum systems we say that because the result of individual measurements 
are not, in general, in a one to one relation with the conditions 
of the experiment. One could say that there are other parameters (known as hidden variables) that
determine the outcome, but thanks to the so called no-go theorems, e.g. 
Bell's theorem\cite{nouvelle}, Kochen-Specker\cite{kochen}, etc., this is unlikely to be the case. 
The most accepted interpretation, at least in the area of quantum information, is that the result of any 
individual measurement is ``created'' in the act of measurement. 
This assumption provides the basis of quantum random number generation, certified using 
Bell-type inequalities\cite{nieto,Bancal}.

We are confronted with two different concepts of randomness, the first one related with
algorithmic information theory that define a random sequence as a incompressible one,
and the second one related with unpredictability of measurements. On one side
they are similar because, if a sequence is incompressible, it has no patterns 
at all and if we can't predict something is because there are no regularities
or patterns that we can use to predict the result of the measurement. 
On the other side the word random is used, in the first case, to describe
a property of sequences of numbers while the second one is applied to single
experiments where each measurement gives us a number.

Can we relate both concepts?
Imagine we make a series of measurements of polarization
of photons, assigning a 1 if we get vertical polarization and a 0 in the other case, and put all
the results together forming a string. Is the final sequence compressible?
It will depend on the initial state the photons were prepared.
If we prepare all of them in a similar vertical polarization  $\ket{1}$, we can predict 
that we will measure  a sequence like ``11111...1'' that is very compressible.
If we prepare the photons in the state $\frac{1}{10}\ket{0} + \frac{\sqrt{99}}{10}\ket{1}$, even though
we don't know the result of any measurement in advance, we can anticipate that the outputs will form a 
compressible sequence. Suppose we get something like:
$$
``1111111011111111111011111111011111111111111111111111101111111010..."
$$
where the ratio between zeroes and ones is $\frac{1}{99}$. We can compress
this sequence by writing the number of 1 instead of 1 themselves, like this:
$$
``70110802407010..."
$$
We can reconstruct the original sequence just by replacing a number between zeroes
by its equivalent in ones, and to perform this translation doesn't need too much code.

Due to the fact we can compress sequences with more ones than zeroes (or more zeroes than ones),
both must appear the same number of times.  Therefore the most incompressible sequence must have the form:
$$\frac{\ket{0}+e^{i\theta}\ket{1}}{\sqrt{2}}.$$

In this case there is no obvious way to compress the sequence. While
there is no algorithm able to test the compressibility, we can use another
condition related to compressibility. From the examples above it is clear that
we need equal number of zeroes and ones; but suppose you split the sequences
every two symbols:
$$
``11 ~10~ 11~ 01~ 11~ 11~ 11~ 01~ 11 01~ ..."
$$

Employing the same arguments of compressibility, we must ask for this number to be random 
that all two-symbol subsequences appear the same number of times.
Let $N(x,y)$ be the number of occurrences of the substring $y$ in the string $x$
then we need:
$$
N(x,``0")\approx N(x,``1")\approx L \frac{1}{2},
$$
where L is the length in bits of the sequence; and
$$
N(x,``00")\approx N(x,``01")\approx N(x,``10")\approx N(x,``11")\approx \frac{L}{2}\frac{1}{4},
$$
as we have $L/2$ subsequences of two symbols, each one with probability $\frac 1 4$. The general expression is:
$$
N(x,y)\approx \frac{L}{n}\frac{1}{2^n},
$$
where n is the length in bits of $``y"$. And therefore:
$$
 \left| \frac{N(x,y)}{L/n} - \frac{1}{2^n} \right| < \epsilon 
$$
where $\epsilon$ is a ``small'' number.

C. Calude showed in \cite{art:calude-borel} the relation between 
the previous ideas and incompressibility. The proposal is to select
$$
\epsilon=\sqrt{\frac{log \,L}{L}},
$$
where again $L$ is the length of the sequence, and check the above inequality
to all subsequences shorter, in bits, than $log (log (L))$. If for all subsequences
the inequality is satisfied we will 
say the sequence is Borel normal\footnote{
	This definition is related with the normal numbers E. Borel defined.}.
In \cite{art:calude-borel} is proved that almost all incompressible sequences
are Borel normal. 

Borel normality is a necessary, but not sufficient condition, to incompressibility. An example of a Borel normal 
sequence that can be compressed is :
$$
0100011011000001010011100101110111...
$$
which is called the Champernowne's constant, built with all binary numbers put in order and forming a single string (0, 1, 00, 01, 10, 11, ...). It was the first example of a Borel normal number, later is was realized that almost all numbers are normal numbers.

Even though we cannot compare effectively the relation between unpredictability
and incompressibility or randomness, we can compare Borel normality and unpredictability in
a sequence of individual measurement outcomes.  

The concepts of Borel normality and incomputablity inspired a comparison between
quantum and computer generated random sequences. While quantum randomness can be
proven incomputable; that is, it is not exactly reproducible by any algorithm, software-generated random numbers, known as pseudo-random, can be reproduced if the computer code and the seed are known.
Calude et al. \cite{art:Calude:ExpEvi} performed finite tests of randomness inspired by algorithmic information theory. They report that all tests produced evidence -with different degrees of statistical significance- of differences between quantum and non-quantum sources. But in a recent investigation of  randomness in ten sequences of bits generated from the differences 
between detection times of photon pairs generated by spontaneous parametric downconversion, they are found fulfilling the randomness criteria without difficulties \cite{Sol15}. A deeper study is being carried out to clarify this point.

\section{Computability and undecidability}

We have explored the close relationship between quantum physics and randomness, 
the definition of random as the absence of a pattern, the association of a pattern with an algorithm,
and, through algorithmic information theory, with Borel normality, arriving at a necessary condition 
for a string to be random which has been employed to analyze the randomness of sequences of photon
detections.   

We can now go a step forward, and explore the relation between physical laws, algorithms and their
computability. As algorithms can be mapped to Turing machines, programs built with finite alphabets, they are infinitely many but numerables.
The set of rational numbers is also infinite numerable, but the real numbers are non-numerable infinite. It follows that the absolute majority of irrational numbers are incomputable in principle: it is impossible to write a computer code to evaluate any of them. It is also impossible to name them, even trying to write a full encyclopedia devoted to each of them is worthless. This concept if thrilling for some of us. It invites to question the meaning of the physical theories, which associate physical observables with real numbers. But, has it any practical implication, or is just a curious but irrelevant information?

A physical law can be seen as a pattern, a regularity which can be described with a mathematical equation relating measurable quantities. It is assumed that the equation can be solved, given a set of initial or boundary conditions, and the solutions provide predictions for the future behavior of the system. 
Any practitioner has found the practical difficulties in doing so. In many cases the dimensions of the basis in which the solution must be expressed are huge, exceeding the ever growing capacities of the supercomputers. In other cases the dynamical equations are chaotic, i.e. display extreme sensitivity to the initial conditions, which puts a limit in the predictive power independent of the computers. We are used to these limitations and have found ways to deal with them, estimating the uncertainties in the predictions and looking for more powerful methods to reduce them.

But there are some mathematical equations which do not have any algorithm which can be used to solve them. 
A beautiful example is given by the following problem in number theory:
Does the equation  $\sum a_{ijk...} x^i y^j z^k...= 0, x_i,a_i \in \mathbb{Z} $, which is
a polynomial equation with integer coefficients, have roots in the integers?
It has been proved in \cite{davis} that there is no algorithm to answer this question.
It is incomputable. No computer can answer it.

An interesting connection of this problem with quantum physics is offered 
in \cite{art:Kieu_hyper}. It is suggested to 
 evaluate the square of the left hand side of the previous equation,
which will then be always equal or greater than zero.
It can be interpreted  as a Hamiltonian, where the integer numbers are associated with photon numbers.
If a physical realization of this Hamiltonian can be constructed, measuring the ground state energy of
the system would provide an answer, if it is zero, to the integer values of  $x,y,z,..$ which are the solutions,
proving their existence. 
There are many subtle points in this proposal,
which can be explored in \cite{art:Kieu_hyper}. 

Are there relevant problems in physics  belonging to this category? 
Examples of undecidability and incompleteness results in classical analysis within a sufficiently rich axiom system
can be extended to undecidability and incompleteness results in axiomatized formulations of physics. Many undecidability and incompleteness results deal with the integrability of Hamilton-Jacobi equations, and with the possibility of proving that a given dynamical system has a chaotic behavior \cite{da1991undecidability}. 

Consider the following question: Given a quantum many-body Hamiltonian, is the system it describes gapped or gapless? 
It is perhaps the simplest and most relevant question in condensed matter, and can be rephrased into the mass problem in quantum field theory.
For a finite number of components (like spins in a Heisenberg chain), the solution can be numerically found. The challenge is to find the thermodynamic limit, when the number of components tends to infinity. 
It has recently been shown that ``this problem is undecidable, in the
same sense as the Halting Problem was proven to be undecidable by Turing. A consequence of this is that the spectral gap of certain
quantum many-body Hamiltonians is not determined by the axioms of mathematics, much as G¨odels incompleteness theorem implies
that certain theorems are mathematically unprovable."\cite{Cub15}

This is a really interesting situation. One is invited to consider the possibility to solve problems
which are demonstrated to be incomputable, for whom it has been proved the impossibility to compute even the
existence of a solution, by employing physical systems which exactly map the problem into an experimental setting.
Measuring the output would provide the answer. In a way it would be a return to analogous computation. In fact, the digital
computation we employ daily also employs programs which are written in bits, each having a physical realization in the computer.
The difference is that in the usual case we could, in principle, perform by hand the same computation. 
To solve incomputable problems  using devices which are not equivalent to Universal Turing Machines is the proposal of a 
research area called  hypercomputation \cite{hyper}. The possibility of effectively employing it in solving problems seems at the present
very far away.

We can go further an apply this kind of ideas to measurement itself. The 
``physical variables'' that we use are always considered as results of procedures
 that we can perform to measure them. Are this procedures algorithmic? 
We have algorithmic procedures to measure many physical quantities like distance or voltage (with some limitations).
The creation or invention of the procedure is not algorithmic but, once it is written, every trained person having the proper equipment could perform it following a list of instructions. 
Can we imagine a ``physical variable'' whose process of measurement is not algorithmic?
If we have a variable whose process is not algorithmic there should be no fixed list of instructions to measure it. This may seem unreal but take a Turing machine and suppose we want to measure if this machine will halt or not,  there is no algorithm to do that, the only option to know the answer is to seat and wait.

\section{Conclusion}

The relation between physics and randomness is indeed rich, being randomness
quite complicated by itself. Being considered in the past a curiosity or a metaphysical question outside of
physics, the developments in the last decades in quantum optics an quantum information, in particular 
 the applications of individual measurements of quantum systems to random number generation and 
security, have brought them to a central stage.

There are very interesting questions about the meaning in physics of concepts
like law, pattern, and their computability. These questions are not only related
with the results of measurements but with some central elements of physics like 
``physical variables''.

\section*{Acknowledgements}

We are very grateful for the contributions of Bogdan Mielnik to illuminate the dark side of quantum mechanics.
This work was supported in part by Conacyt, Mexico. AS is a fellow of Conacyt, Mexico.

\section*{References}
\bibliography{./bibliografia.bib}

\providecommand{\newblock}{}
\begin{thebibliography}{10}
\expandafter\ifx\csname url\endcsname\relax
  \def\url#1{{\tt #1}}\fi
\expandafter\ifx\csname urlprefix\endcsname\relax\def\urlprefix{URL }\fi
\providecommand{\eprint}[2][]{\url{#2}}

\bibitem{bell}
 2014 {\em Journal of Physics A: Mathematical and Theoretical\/} {\bf 47}
  number 42

\bibitem{wiseman14}
Wiseman H 2014 {\em Nature.\/} {\bf 510} 467--469

\bibitem{Zei99}
Zeilinger A 1999 {\em Foundations of Physics,\/} {\bf 29} 631--643

\bibitem{Sca10}
Scarani V 2010 {\em Nature\/} {\bf 464} 988--989

\bibitem{Pir10}
Pironio~et al S 2010 {\em Nature\/} {\bf 464} 1021--1024

\bibitem{chris13}
Christensen B~G, McCusker K~T, Altepeter J~B, Calkins B, Gerrits T, Lita A~E,
  Miller A, Shalm L~K, Zhang Y, Nam S~W, Brunner N, Lim C~C~W, Gisin N and
  Kwiat P~G 2013 {\em Phys. Rev. Lett.\/} {\bf 111} 130406

\bibitem{bancal14}
Bancal J~D, Sheridan L and Scarani V 2014 {\em New J. Phys.\/} {\bf 16} 033011

\bibitem{nieto14}
Nieto-Silleras O, Pironio S and Silman J 2014 {\em New J. Phys.\/} {\bf 16}
  013035

\bibitem{chaitin}
Chaitin G 2006 {\em Scientific American\/} {\bf March} 74--81

\bibitem{art:turing}
Turing A 1937 {\em Proceedings of the London Mathematical Society\/} {\bf 42}
  230--265

\bibitem{davis}
Davis M 1953 {\em Computability and Unsolvability\/} (Dover)

\bibitem{book:Calude}
Calude C 2002 {\em Information and Randomness: An Algorithmic Perspective\/}
  2nd ed (Secaucus, NJ, USA: Springer-Verlag New York, Inc.) ISBN 3540434666

\bibitem{book:Li}
Li M and Vitnyi P~M 2008 {\em An Introduction to Kolmogorov Complexity and Its
  Applications\/} 3rd ed (Springer Publishing Company, Incorporated) ISBN
  0387339981, 9780387339986

\bibitem{pi}
Lopez-Ortiz A 1998 How to compute digits of pi?
  \urlprefix\url{https://cs.uwaterloo.ca/{\raise.17ex\hbox{$\scriptstyle\sim$}}alopez-o/
  math-faq/mathtext/node12.html}

\bibitem{art:Zei05}
Zeilinger A 2005 {\em Nature\/} {\bf 438} 743

\bibitem{nouvelle}
Bell J~S and Aspect A 2004 {\em Speakable and Unspeakable in Quantum
  Mechanics\/} (Cambridge University Press) pp 232--248 2nd ed ISBN
  9780511815676 cambridge Books Online
  \urlprefix\url{http://dx.doi.org/10.1017/CBO9780511815676.026}

\bibitem{kochen}
Kochen S and Specker E 1975 {\em The Logico-Algebraic Approach to Quantum
  Mechanics\/} ({\em The University of Western Ontario Series in Philosophy of
  Science\/} vol~5a) ed Hooker C (Springer Netherlands) pp 293--328 ISBN
  978-90-277-0613-3
  \urlprefix\url{http://dx.doi.org/10.1007/978-94-010-1795-4\_17}

\bibitem{nieto}
Nieto-Silleras O, Pironio S and Silman J 2014 {\em New Journal of Physics\/}
  {\bf 16} 013035
  \urlprefix\url{http://stacks.iop.org/1367-2630/16/i=1/a=013035}

\bibitem{Bancal}
Bancal J~D, Sheridan L and Scarani V 2014 {\em New Journal of Physics\/} {\bf
  16} 033011 \urlprefix\url{http://stacks.iop.org/1367-2630/16/i=3/a=033011}

\bibitem{art:calude-borel}
Calude C 1994 {\em in G. Rozenberg, A. Salomaa (eds.) {\it Developments in
  Language Theory}\/}  113--129

\bibitem{art:Calude:ExpEvi}
Calude C~S, Dinneen M~J, Dumitrescu M and Svozil K 2010 {\em Phys. Rev. A\/}
  {\bf 82}(2) 022102
  \urlprefix\url{http://link.aps.org/doi/10.1103/PhysRevA.82.022102}

\bibitem{Sol15}
Solis A, Angulo~Mart{\'\i}nez A~M, Ram{\'\i}rez-Alarc\'on R, Cruz-Ram{\'\i}rez
  H, U'Ren A~B and Hirsch J~G 2015 {\em Physica Scripta\/}  in press

\bibitem{art:Kieu_hyper}
Kieu T~D 2003 {\em Int.J.Theor.Phys.\/} {\bf 42} 1461--1478

\bibitem{da1991undecidability}
da~Costa N~C and Doria F~A 1991 {\em International Journal of Theoretical
  Physics\/} {\bf 30} 1041--1073

\bibitem{Cub15}
Cubitt T, Perez-Garcia D and Wolf M~M 2015 {\em arXiv\/} {\bf [quant-ph]}
  1502.04135

\bibitem{hyper}
Copeland B~J and Proudfoot D 1999 {\em Scientific American\/} {\bf April}
  98--103

\end{thebibliography}

\end{document}